\documentclass[twoside,11pt]{article}

\usepackage[abbrvbib,preprint]{jmlr2e}
\usepackage[frozencache]{minted}
\usepackage{textcomp}
\usepackage{graphicx}
\usepackage{amsmath}
\usepackage{bm}
\usepackage{booktabs}      
\usepackage{multirow}      
\usepackage{tcolorbox}   
\usepackage{parcolumns}
\usepackage{adjustbox}
\usepackage{nicefrac}
\usepackage{tabularx}
\usepackage{array}
\usepackage{wasysym}
\usepackage{dcolumn}

\hypersetup{pdfborderstyle={/S/S/W 1},colorlinks=true,linkcolor=blue,citecolor=blue,filecolor=black,urlcolor=blue}
\hypersetup{pdfborder={0 0 0},citebordercolor={0 0 0},filebordercolor={0 0 0},linkbordercolor={0 0 0},menubordercolor={0 0 0},urlbordercolor={0 0 0}}
\hypersetup{pdftitle={The ensmallen library for flexible numerical optimization}}
\hypersetup{pdfauthor={Ryan R. Curtin, Marcus Edel, Rahul Ganesh Prabhu, Suryoday Basak, Zhihao Lou, Conrad Sanderson}}

\newcolumntype{d}[1]{D{.}{.}{#1}}

\newcolumntype{R}[2]{%
  >{\adjustbox{angle=#1,lap=\width-(#2)}\bgroup}%
  l%
  <{\egroup}%
}

\usepackage{lastpage}

\firstpageno{1}

\makeatletter
\renewcommand\@makefntext[1]{%
\setlength\parindent{1em}%
\noindent
{#1}}
\makeatother

\makeatletter
\def\blfootnote{\xdef\@thefnmark{}\@footnotetext}
\makeatother

\begin{document}

\title{The ensmallen library for flexible numerical optimization}

\author{%
  \name Ryan R. Curtin \hfill \email ryan@ratml.org ~|~ \addr RelationalAI, Atlanta, GA, USA\\
  \name Marcus Edel \hfill \addr Free University of Berlin, Germany\\
  \name Rahul Ganesh Prabhu \hfill \addr Birla Institute of Technology and Science Pilani, India\\
  \name Suryoday Basak \hfill \addr University of Texas at Arlington, USA\\
  \name Zhihao Lou \hfill \addr Epsilon, Chicago, IL, USA\\
  \name Conrad Sanderson \hfill \addr Data61/CSIRO, Australia, and Griffith University, Australia%
  }

\editor{~}

\maketitle

\begin{abstract}%
We overview the \textit{ensmallen} numerical optimization library,
which provides a flexible C++ framework
for mathematical optimization of user-supplied objective functions.
Many types of objective functions are supported,
including general, differentiable, separable, constrained, and categorical.
A~diverse set of pre-built optimizers is provided,
including Quasi-Newton optimizers and many variants of Stochastic Gradient Descent.
The underlying framework facilitates the implementation of new optimizers.
Optimization of an objective function typically requires supplying only one or two {C++} functions.
Custom behavior can be easily specified via callback functions.
Empirical comparisons show that \textit{ensmallen}
outperforms other frameworks while providing more functionality.
The library is available at {\footnotesize \url{https://ensmallen.org}}
and is distributed under the permissive BSD license.

\blfootnote{\textbf{Published in:} Journal of Machine Learning Research, Vol. 22, No. 166, 2021.\\ \href{https://jmlr.org/papers/v22/20-416.html}{https://jmlr.org/papers/v22/20-416.html}}

\vspace{1ex}

\end{abstract}

\begin{keywords}
  Numerical optimization, mathematical optimization, function minimization.
\end{keywords}

\vspace{-1ex}
\section{Introduction}

The problem of numerical optimization is generally expressed as
$\operatornamewithlimits{argmin}_x f(x)$
where $f(x)$ is a given objective function and $x$ is typically a vector or matrix.
Such optimization problems are fundamental and ubiquitous in the computational sciences~\citep{Nocedal_2006}.
Many frameworks or libraries for specific machine learning approaches
have an integrated optimization component for distinct and limited use cases,
such as
TensorFlow~\citep{TensorFlow_arXiv_2016},
PyTorch~\citep{PyTorch_NeurIPS_2019}
and LibSVM~\citep{libsvm2011}.
There are also many general numerical optimization toolkits
aimed at supporting a wider range of use cases,
including SciPy~\citep{SciPy_NMeth_2020},
opt++~\citep{meza1994opt++},
and 
OR-Tools~\citep{ortools} among many others.
However, such toolkits still have limitations in several areas,
including:
(i)~types of supported objective functions,
(ii)~selection of available optimizers,
(iii)~support for custom behavior via callback functions,
(iv)~support for various underlying element and matrix types used by objective functions,
and
(v)~extensibility, to facilitate adding more optimizers.

These shortcomings have motivated us to create the \textit{ensmallen} library,
which explicitly supports numerous types of user-defined objective functions,
including general, differentiable, separable, categorical, and constrained
objective functions, as well as semidefinite programs.
Custom behavior during optimization can be specified via {callback} functions,
for purposes such as printing progress, early stopping, inspection and modification of an optimizer's state,
and debugging of new optimizers.
A~large set of pre-built optimizers is provided;
at the time of writing, 46 optimizers are available.
This includes 
simulated annealing \citep{kirkpatrick1983optimization},
several Quasi-Newton optimizers \citep{liu1989limited,mokhtari2018},
and many variants of Stochastic Gradient Descent \citep{Ruder_2016}.

The user interface to the optimizers is intuitive
and matches the ease of use of popular
optimization toolkits mentioned above; for more details, see
the online documentation at
\mbox{\small\url{https://ensmallen.org/docs.html}}.
Typically, a user only needs to implement one or two {C++}
functions, and then they can use any optimizer
matching the type of their objective.

Importantly, the ease-of-use does not come at the cost of efficiency; instead,
\textit{ensmallen} uses C++ template metaprogramming techniques (hidden from the
user) to provide accelerations and simplifications where possible.
The use of various underlying element and matrix types is supported,
including single- and double-precision floating point,
integer values, and sparse data.
Lastly, \textit{ensmallen} provides an extensible framework to easily allow the implementation of new optimization techniques.

\vspace{-1ex}
\section{Functionality}

The task of optimizing an objective function with \textit{ensmallen} is straightforward.
The type of objective function defines the implementation requirements.
Each type has a minimal set of methods that must be implemented;
typically between one and four methods.
Apart from the requirement of an implementation of $f(x)$,
characteristics of $f(x)$ can be exploited through additional functions.
For example, if $f(x)$ is differentiable,
an implementation of $f'(x)$ can be used to accelerate the optimization process.
Then, one of the pre-built differentiable function optimizers,
such as L-BFGS~\citep{liu1989limited}, can be used.

Whenever possible, \textit{ensmallen} will automatically infer methods
that are not provided.
For example, given a separable objective function
$f(x) = \sum_i f_i(x)$
where an implementation of $f_i(x)$ is provided
(as well as the number of such separable objectives),
an implementation of $f(x)$ can be automatically inferred.
This is done at compile-time, and so there is no additional runtime
overhead compared to a manual implementation.
C++ template metaprogramming techniques~\citep{Abrahams_2004,alexandrescu2001modern}
are internally used to automatically produce efficient code during compilation.

To implement a new optimizer, the user only needs to implement a class with an
{\tt Optimize()} method taking an external implementation of $f(x)$ (and other
functions specific to the class of objective function).  As such, \textit{ensmallen}
is easily extensible.

When an optimizer (either pre-built or new) is used with a user-provided objective function,
the requirements for that optimizer are checked
(e.g., presence of an implementation of $f'(x)$),
resulting in user-friendly error messages at compile-time
if there are any issues.
For example, as L-BFGS is suited for differentiable functions,
a compile-time error will be printed if an attempt is made
to use it with non-differentiable (general) functions.

\begin{figure}[t!]
\hrule
\vspace{1ex}
\centering
\begin{minted}[fontsize=\footnotesize]{c++}

#include <ensmallen.hpp>

struct LinearRegressionFn
{
  LinearRegressionFn(const arma::mat& in_X, const arma::vec& in_Y) : X(in_X), y(in_Y) {}

  double Evaluate(const arma::mat& phi)
    { const arma::vec tmp = X.t() * phi - y;  return arma::dot(tmp, tmp); }
  
  void Gradient(const arma::mat& phi, arma::mat& grad)
    { grad = 2 * X * (X.t() * phi - y); }

  const arma::mat& X; const arma::vec& y;
};

int main() 
{
  arma::mat X; arma::vec y;
  // ... set the contents of X and y here ...
  arma::mat phi_star(X.n_rows, 1, arma::fill::randu);  // initial point (uniform random)
  LinearRegressionFn f(X, y);
  ens::L_BFGS optimizer; // create an optimizer object with default parameters
  optimizer.Optimize(f, phi_star); // after here, phi_star contains the optimized parameters
}
\end{minted}
\hrule
\vspace*{-0.5em}
\caption{Example implementation of an objective function class for linear
regression and usage of the L-BFGS optimizer.
The optimizer can be easily changed by replacing
{\tt ens::L\_BFGS} with another optimizer,
such as {\tt ens::GradientDescent},
or {\tt ens::SA} which implements simulated annealing \citep{kirkpatrick1983optimization}.
}
\label{fig:lr_function}
\vspace*{-2em}
\end{figure}

\vspace{-1ex}
\section{Example Usage \& Empirical Comparison}

For an example implementation and comparison, let us first consider linear regression.
In this problem, predictors $\bm X \in \mathcal{R}^{d \times n}$
and associated responses $\bm y \in \mathcal{R}^n$ are given.
We wish to find the best linear model $\bm \Phi \in \mathcal{R}^d$,
which translates to finding
$\bm \Phi^* = \operatornamewithlimits{argmin}_{\bm\Phi} f(\bm \Phi)$ for
$f(\bm \Phi) = \| \bm X^{\top} \bm \Phi - \bm y \|^2.$
This gives the gradient
$f'(\bm \Phi) = 2 \bm X (\bm X^{\top} \bm \Phi - \bm y).$

To find $\bm \Phi^*$ using a differentiable optimizer,
we simply need to provide implementations of $f(\bm \Phi)$ and $f'(\bm \Phi)$.
For a differentiable function, \textit{ensmallen} requires only two methods:
{\tt Evaluate()} and {\tt Gradient()}.
The pre-built L-BFGS optimizer can then be used to find~$\bm \Phi^*$.
Figure~\ref{fig:lr_function} shows an example implementation.
Via the use of the Armadillo library~\citep{sanderson_2016,sanderson_2019},
the linear algebra expressions to implement the objective function and its gradient
are compact and closely match natural mathematical notation.
Armadillo efficiently translates the expressions into standard BLAS and LAPACK function calls~\citep{anderson1999lapack,Psarras_2022},
allowing easy exploitation of high-performance implementations such as the multi-threaded \mbox{OpenBLAS}~\citep{OpenBLAS} and Intel MKL~\citep{IntelMKL} libraries.

Table~\ref{tab:lbfgs} compares the performance
of \textit{ensmallen} against other frameworks
for the linear regression problem on various dataset sizes.
We compare against {\tt SciPy},
{\tt Optim.jl} \citep{mogensen2018optim},
and the {\tt bfgsmin()} function from GNU Octave \citep{octave}.
We also compare against the automatic differentiation implementations of
PyTorch, TensorFlow,
and the Python library Autograd \citep{maclaurin2015autograd}.
In each framework, the provided L-BFGS optimizer is limited to $10$ iterations.
Highly noisy random data with a slight linear pattern is used.
The runtimes are the average of 5 runs.
The experiments were performed on an AMD Ryzen 7 2700X with 64GB RAM,
with g++ 10.2.0, Julia 1.5.2, Python 3.8.5, and Octave 6.1.0.
For fairness, all tools used the CPU only.

Next, we consider the common machine learning problem of logistic regression
using two-class versions of various real datasets from the UCI dataset
repository~\citep{ucimlrepository}.  The setup of our experiments is the same as for
the previous example; results are in
Table~\ref{tab:lbfgs_logistic_regression}.

Both simulations show that \textit{ensmallen} achieves the lowest runtimes,
sometimes by large margins.  This is due to multiple factors, including the
efficiency of the optimizer implementations in \textit{ensmallen},
template metaprogramming optimizations in Armadillo and \textit{ensmallen}, and
minimal overhead and dependencies compared to the competitors.

\begin{table}[b!]
{\small
\centering
\begin{tabular}{l *{5}{d{3.4}}}
\toprule
{\em Framework} & \multicolumn{1}{c}{$d$: 100, $n$: 1k} & \multicolumn{1}{c}{$d$: 100, $n$: 10k} & \multicolumn{1}{c}{$d$: 100, $n$: 100k} & \multicolumn{1}{c}{$d$: 1k, $n$: 100k} \\
\midrule
\texttt{ensmallen}  & {\bf 0}.{\bf 0016}s & {\bf 0}.{\bf 0067}s & {\bf 0}.{\bf 1460}s & {\bf 1}.{\bf 4011s} \\
\texttt{Optim.jl}   & 0.0069s       & 0.0117s       & 0.1672s       & 1.3985s \\
\texttt{SciPy}      & 0.0028s       & 0.0110s       & 0.2247s       & 1.8461s \\
Autograd            & 0.0073s       & 0.0163s       & 0.2416s       & 1.8733s \\
PyTorch             & 0.0469s       & 0.0986s       & 0.5670s       & 5.6041s \\
TensorFlow          & 0.1876s       & 0.2306s       & 0.6925s       & 6.6764s \\
\texttt{bfgsmin()}  & 1.9773s       & 18.0515s      & 123.437s      & 9710.6750s \\
\bottomrule
\end{tabular}
\vspace*{-0.7em}
\caption{
Runtimes for optimizing linear regression parameters on various dataset sizes,
where $n$ is the number of samples,
and $d$ is the dimensionality of each sample.
}
\label{tab:lbfgs}
}
\vspace*{-1.2em}
\end{table}

\begin{table}[b!]
{\small
\centering
\begin{tabular}{l *{5}{d{6.5}}}
\toprule
{\em Framework} & \multicolumn{1}{c}{MNIST} & \multicolumn{1}{c}{covertype} & \multicolumn{1}{c}{pokerhand} & \multicolumn{1}{c}{font} & \multicolumn{1}{c}{isolet} \\
  & \multicolumn{1}{c}{60k $\times$ 784} & \multicolumn{1}{c}{407k $\times$ 55} & \multicolumn{1}{c}{700k $\times$ 10} & \multicolumn{1}{c}{832k $\times$ 407} & \multicolumn{1}{c}{7.8k $\times$ 617}
\\
\midrule
\texttt{ensmallen}  & {\bf 0}.{\bf 6546s} & {\bf 0}.{\bf 9038s} & {\bf 0}.{\bf 5186s} & {\bf 6}.{\bf 1678s} & {\bf 0}.{\bf 0510s} \\
\texttt{Optim.jl}   & 1.4231s       & 1.2067s       & 0.6754s       & 10.9051s      & 0.1214s \\
\texttt{SciPy}      & 0.8101s       & 1.1388s       & 1.0231s       & 7.5838s       & 0.07519s \\
Autograd            & 0.8012s       & 1.4241s       & 2.6005s       & 7.1224s       & 0.0876s \\
PyTorch             & 6.5710s       & 8.8340s       & 3.2404s       & 59.0194s      & 0.8172s \\
TensorFlow          & 9.3662s       & 5.4231s       & 2.6005s       & 70.1122s      & 0.7563s \\
\texttt{bfgsmin()}  & 539.1358s     & 43.9067s      & 8.2561s       & 2358.1680s    & 48.8020s \\
\bottomrule
\end{tabular}
\vspace*{-0.7em}
\caption{
Runtimes for training a logistic regression model on
real data with L-BFGS.}
\label{tab:lbfgs_logistic_regression}
}
\vspace*{-1.2em}
\end{table}

\vspace{-1ex}
\section{Conclusion}

The \textit{ensmallen} numerical optimization provides a flexible framework
for optimization of user-supplied objective functions in C++.
Unlike other frameworks, \textit{ensmallen} supports many types of objective functions,
provides a diverse set of pre-built optimizers,
supports custom behavior via callback functions,
and handles various element and matrix types used by objective functions.
The underlying framework facilitates the implementation of new optimization techniques,
which can be contributed for inclusion into the library.

The library has been successfully used by open source projects
such as the {\it mlpack} machine learning toolkit~\citep{mlpack4_2023}.
The library uses the permissive BSD license~\citep{Laurent_2008},
with the development done in an open and collaborative manner.
The source code and documentation are freely available at \mbox{\url{https://ensmallen.org}}.

Further details, such as internal use of template metaprogramming
for automatic generation of efficient code, automatic function inference,
clean error reporting, and various approaches for obtaining efficiency
are all discussed in the accompanying technical report~\citep{ensmallen2020}.

\small
\bibliography{refs}

\end{document}